\documentclass[a4paper,fleqn]{cas-sc}

\usepackage[numbers]{natbib}
\usepackage[ruled,vlined,linesnumbered]{algorithm2e}

\def\tsc#1{\csdef{#1}{\textsc{\lowercase{#1}}\xspace}}
\tsc{WGM}
\tsc{QE}

\begin{document}
\let\WriteBookmarks\relax
\def\floatpagepagefraction{1}
\def\textpagefraction{.001}

\shorttitle{}    

\shortauthors{}  

\title [mode = title]{ScaleAcross: Designing Multi-Data-Center Infrastructure for Geo-Distributed AI Training}  



%

\author[1]{Naved Inam}

\ead{navedw2469@gmail.com}

\credit{Methodology, Software, Investigation, Writing -- original draft}

\affiliation[1]{organization={Department of Computer Science and Engineering, Indian Institute of Technology Indore},
            city={Indore},
            postcode={453552},
            state={Madhya Pradesh},
            country={India}}

\author[1]{Aryan Alpesh Bhavsar}

\ead{phd2401201001@iiti.ac.in}

\credit{Software, Validation, Data curation}

\author[1]{Masabattula Teja Nikhil}

\ead{ms2404101014@iiti.ac.in}

\credit{Formal analysis, Validation, Investigation}

\author[2]{Sidharth Sharma}[orcid=0000-0003-0344-4937]
\cormark[1]

\ead{sidharth@iitj.ac.in}

\credit{Conceptualization, Supervision, Methodology, Writing -- review \& editing}

\affiliation[2]{organization={Department of Computer Science and Engineering, Indian Institute of Technology Jodhpur},
            city={Jodhpur},
            postcode={342030},
            state={Rajasthan},
            country={India}}

\cortext[1]{Corresponding author}




\begin{abstract}
The rapid growth of AI models and increasing data sovereignty requirements are driving the transition toward geo-distributed AI training across multiple data centers. Such deployments introduce system-level challenges arising from synchronization-intensive communication, cross-site data exchange, and wide-area latency constraints. This paper investigates EVPN--VXLAN as an infrastructure foundation for geo-distributed AI training environments and presents a scalable emulation framework for systematically studying distributed AI workloads under realistic wide-area conditions. The proposed framework combines VXLAN overlays with EVPN-based inter-data-center connectivity and is implemented using ContainerLab and FRRouting (FRR). The framework further incorporates Equal-Cost Multi-Path (ECMP) routing, Bidirectional Forwarding Detection (BFD), and a queue-pair-aware traffic distribution mechanism designed to improve communication behavior for synchronization-intensive AI workloads while preserving compatibility with commodity infrastructure. Using realistic WAN emulation, we characterize communication and system behavior under distributed training workloads employing AllReduce and Parameter Server communication patterns. Results provide insights into traffic distribution, resilience, and infrastructure behavior in geo-distributed AI environments, highlighting the potential of reproducible multi-data-center infrastructure frameworks for scalable distributed AI training.
\end{abstract}




\maketitle

\section{Introduction}

The rapid growth of Deep Neural Networks (DNNs), particularly Large Language Models (LLMs), has pushed single data-center infrastructure toward limits in compute density, power delivery, and thermal capacity \cite{dong2025beyond, evolution-of-ai-dc, pannocchi2024datacenter}. Consequently, AI organizations are increasingly exploring collaborative training across geographically distributed data centers to improve scalability, resilience, resource utilization, and compliance with emerging data sovereignty requirements \cite{nvidia2024turbocharge}. Such deployments allow data to remain within jurisdictional boundaries while enabling geographically distributed compute resources to participate in collaborative training through parameter or gradient exchange.

However, geo-distributed AI training introduces system-level challenges beyond simply scaling GPU resources across multiple locations. Modern distributed training relies heavily on synchronization-intensive communication patterns, where collective operations such as AllReduce are highly sensitive to latency and bandwidth constraints \cite{gao2025scc}. Inter-region delays can significantly slow parameter synchronization, making communication infrastructure a critical component of overall training efficiency. Furthermore, geo-distributed deployments require orchestration frameworks to maintain seamless connectivity, service discovery, workload mobility, and policy management across geographically separated clusters \cite{201564, mcmahan2023communicationefficientlearningdeepnetworks, 10.1145/3448016.3452773, L3DML}. These requirements elevate communication and system infrastructure design into a key consideration for scalable geo-distributed AI training.

Existing transport technologies, such as RoCEv2 \cite{rdma-meta} and emerging AI-oriented fabrics such as Ultra Ethernet (UEC) \cite{hoefler2025ultra}, primarily optimize communication within tightly coupled AI infrastructure. RoCEv2 enables efficient RDMA communication inside AI clusters but does not naturally extend across wide-area deployments \cite{rdma-hyperscale}. Similarly, UEC emphasizes high-throughput communication and congestion management for short- to medium-reach AI infrastructure. While these technologies remain essential for intra-data-center communication, geo-distributed AI training additionally requires scalable inter-data-center connectivity, workload mobility, tenant isolation, and resilient infrastructure management across geographically dispersed sites.

VXLAN \cite{rfc7348} combined with EVPN \cite{rfc8365} provides a promising foundation for addressing these requirements. VXLAN enables scalable Layer-2 overlays across Layer-3 infrastructure, while EVPN provides endpoint discovery, routing scalability, and operational resilience through a standards-based control plane. Together, they enable geographically distributed AI infrastructure to operate over a unified virtualized communication fabric while preserving operational flexibility and multi-tenancy.

Despite widespread adoption in cloud environments, the applicability of EVPN and VXLAN to geo-distributed AI training remains relatively unexplored. AI training workloads exhibit unique communication characteristics, including bursty synchronized collective traffic, long-lived elephant flows, and sensitivity to transient failures and path imbalance. Existing Equal-Cost Multi-Path (ECMP) routing mechanisms may exhibit poor entropy under synchronized training traffic patterns, resulting in path collisions and uneven utilization \cite{rdma-meta}.

This work investigates the applicability of EVPN and VXLAN for geo-distributed AI training environments and presents a reproducible emulation framework for systematically studying distributed AI workloads under realistic WAN conditions. We further introduce network-level enhancements to improve resilience and communication efficiency while preserving compatibility with commodity infrastructure. Building and operating geo-distributed AI infrastructure is both expensive and operationally complex; therefore, the proposed framework provides a practical platform for exploring networking and system design choices prior to real-world deployment.

The main contributions are summarized below:

\begin{enumerate}

\item We design an EVPN--VXLAN-based emulated communication fabric tailored for geo-distributed AI training workloads and provide an open, reproducible ContainerLab-based framework for evaluating distributed AI training over WAN environments.

\item We propose a queue-pair-aware source-port allocation mechanism that improves ECMP path diversity for distributed AI training workloads without modifying existing ECMP hashing pipelines or packet formats.

\item We experimentally characterize communication behavior and system performance under geo-distributed AI training deployments involving WAN delays, enhanced ECMP mechanisms, and failure recovery scenarios.

\end{enumerate}

\section{Architectural Challenges} \label{sec:arch-challenges}
Training AI models across geo-distributed DCs brings unique challenges.
This section outlines the key architectural hurdles in designing network infrastructure for AI training.


\subsection{Synchronization Overhead in Distributed Training}

In distributed training, the synchronization of the model parameters is essential for convergence. In a parameter server (PS) architecture, workers periodically push their gradients to centralized servers and pull updated parameters for the next iteration. 
Although effective in LAN settings, this approach becomes problematic in WAN environments, where network latencies are higher and bandwidth is more limited.
Even moderate-scale Deep Neural Networks (such as ResNet, Transformer) can involve hundreds of megabytes of model state that must be exchanged every few seconds.

AllReduce-based architectures, commonly used with frameworks like Horovod and NCCL, eliminate the central server by having peers directly exchange gradients. However, these architectures are highly sensitive to WAN-induced jitter and require consistent high-bandwidth links to maintain efficiency. As a result, both approaches face serious bottlenecks in geo-distributed settings unless the underlying communication fabric is optimized to handle inter-site traffic intelligently.

\subsection{Scalability Limitations of Legacy VLAN Architectures}

Traditional VLANs, based on IEEE 802.1Q, support a maximum of 4096 VLAN identifiers, limiting their applicability in modern large-scale and multi-tenant data centers. Contemporary large language model (LLM) training clusters often span tens of thousands of GPU or compute nodes across multiple racks and data centers, far exceeding this limit \cite{jiang2024megascale}. 
Additionally, VLANs operate strictly at Layer 2, which complicates routing across physical sites or availability zones. Maintaining large MAC address tables and broadcast domains further leads to flooding, loops, and convergence delays, making VLAN-based setups unsuitable for the dynamic and distributed nature of AI and ML training workloads.

\subsection{Fault Tolerance and Network Convergence Time}
During AI training, the ability to quickly recover from node or link failures is critical \cite{jiang2024megascale}. Traditional routing protocols such as OSPF or BGP can take several seconds to detect and recover from such events. In contrast, ML training pipelines expect fast failover to prevent wasted GPU cycles and inconsistent training states. Without sub-second convergence, failures during synchronization steps can lead to degraded performance or even training collapse \cite{jiang2024megascale}.

\subsection{Multi-Tenancy and Traffic Isolation}
In multi-tenant environments where multiple users or applications train DNN models concurrently, isolation of traffic and compute resources is essential. The system must support thousands of simultaneous virtual networks, with isolation and minimal interference. Overlay networking is typically used to solve this, but it must be scalable, resilient, and compatible with modern control plane technologies.

\section{VXLAN-EVPN and Proposed Add-ons} \label{sec:architecture}
To address the above challenges, this work adopts a protocol stack composed multiple technologies, each selected based on its ability to meet the specific performance, scalability, and resiliency requirements of geo-distributed AI training systems.

\subsection{VXLAN for Layer 2 Overlays}
Virtual Extensible LAN (VXLAN) is a Layer 2 overlay tunneling protocol that encapsulates Ethernet frames within UDP packets, allowing Layer 2 connectivity to be extended across Layer 3 networks \cite{rfc7348}. VXLAN supports up to 16 million virtual network identifiers (VNIs), compared to only 4096 in VLANs, making it suitable for large-scale, multi-tenant cloud environments.

In this article, VXLAN is used to connect training nodes and parameter servers (or workers in AllReduce) across emulated data center boundaries, allowing the system to maintain logical proximity while being physically dispersed. By encapsulating traffic, VXLAN isolates tenant traffic and reduces the impact of broadcast and multicast storms.

\subsection{EVPN Control Plane with MP-BGP}
Ethernet VPN (EVPN) provides the control plane for VXLAN overlays, replacing flood-and-learn with BGP-driven route distribution \cite{rfc8365}. Using Multi-Protocol BGP (MP-BGP) \cite{rfc4760}, EVPN advertises MAC/IP reachability, multicast groups, and mobility events through well-defined route types (e.g., Type~2 MAC/IP, Type~3 Inclusive Multicast). This enables automatic Virtual Tunnel EndPoint (VTEP) discovery, dynamic tunnel creation, and propagation of tenant-specific MAC/IP information without manual configuration. The use of Route Distinguishers (RDs) and Route Targets (RTs) further supports multi-tenancy and flexible policy control, making EVPN+MP-BGP highly scalable for distributed AI training overlays.





\begin{figure*}
    \centering
    \includegraphics[width=\linewidth]{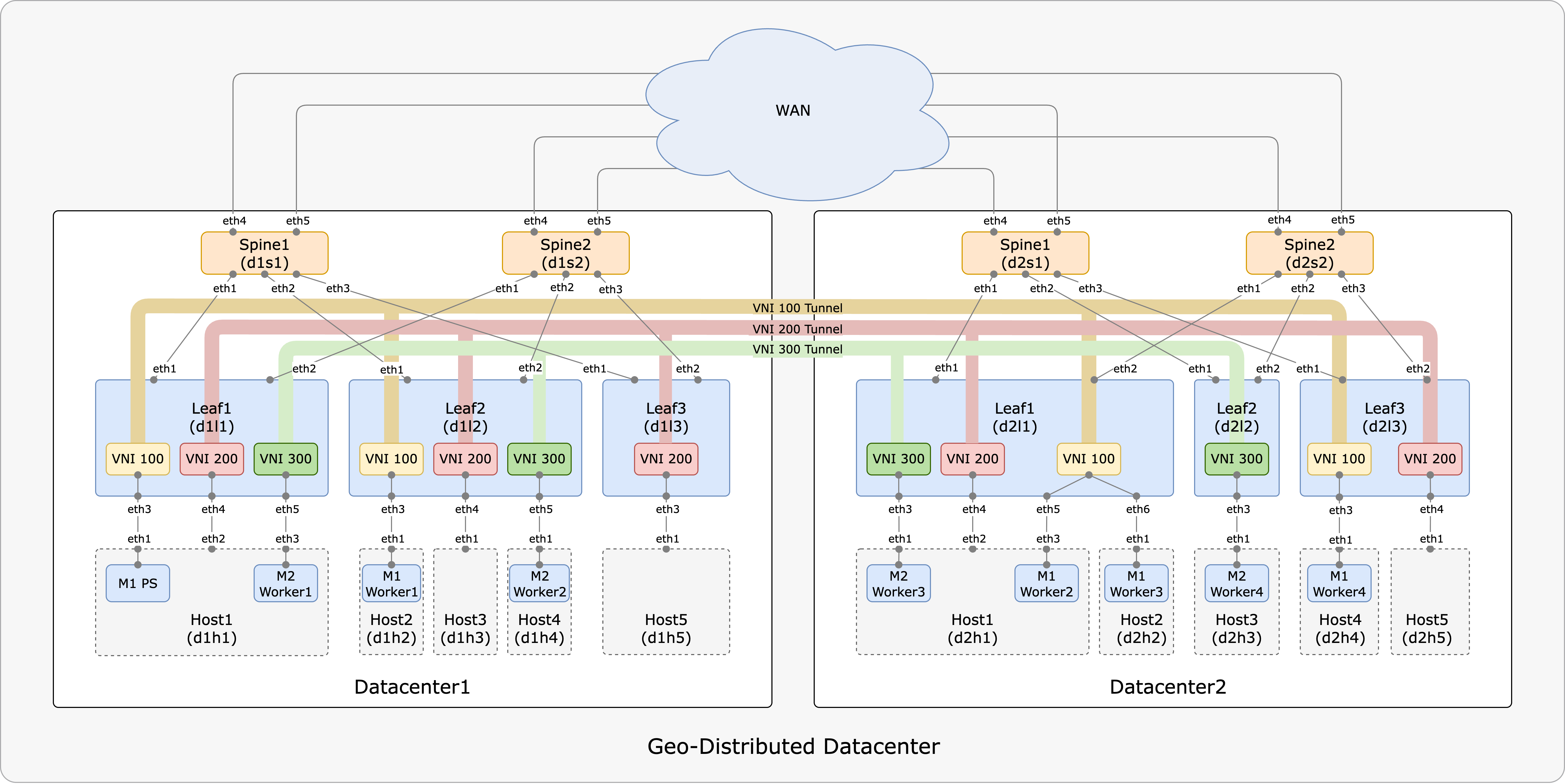}
    \caption{Experimental topology showing simultaneous deployment of Parameter Server (M1) and All Reduce (M2) architectures across the emulated geo-distributed data centers.}
    \label{fig:dist-topology}
\end{figure*}

\subsection{Improving ECMP Path Diversity for AI Training Traffic}

Modern datacenter networks predominantly rely on ECMP routing \cite{hopps2000analysis}, which distributes traffic by hashing packet 5-tuples across multiple equal-cost paths. However, prior work has shown that ECMP performs poorly under distributed training workloads because synchronized communication patterns often produce low flow entropy \cite{al2010hedera,alizadeh2014conga,rdma-meta}. 
Multiple large flows may therefore collide onto identical network paths, creating persistent hotspots and uneven utilization.

Distributed AI training commonly relies on RDMA-based transport layers, where collective communication libraries such as NCCL partition large tensors into multiple chunks that are transmitted concurrently through independent Queue Pairs (QPs) \cite{hu2025demystifying}. 
Each QP forms an independent communication flow participating in collective operations, increasing communication parallelism while simultaneously amplifying the sensitivity of ECMP path selection to flow entropy. 
For example, a 4GB gradient using four channels is divided into four 1GB chunks, where each chunk is assigned to a separate QP and processed concurrently. The network stack treats these QPs as distinct flows identified by unique packet 5-tuples.
RDMA drivers dynamically assign source ports during QP initialization by hashing the QP identifier and mapping the resulting value into the dynamic UDP port range. 
In RoCEv2-based RDMA implementations such as Soft-RoCE (\texttt{rdma-rxe}), source ports are allocated from the range 49,192--65,535. Specifically, the driver hashes the 32-bit QP number to generate a 14-bit offset value, producing a range of $2^{14}=16{,}384$ possible offsets (0--16,383). This offset is then added to the base port value (49,192) to obtain the final source-port assignment.

Although statistically effective, previous work \cite{rdma-meta} identified production scenarios in which different QPs communicating between the same source-destination GPU pair receive identical source ports. This produces identical packet 5-tuples, causing ECMP to map otherwise independent communication channels onto the same physical path and reducing effective traffic distribution.

\subsubsection{Queue-Pair-Aware Source-Port Allocation}

To improve ECMP path diversity without modifying switch hashing logic or packet formats, we redesign source-port assignment inside Soft-RoCE. The proposed mechanism partitions the dynamic UDP source-port range into $k$ non-overlapping bins ($k=4$ in our implementation). Since Soft-RoCE exposes 16,384 possible source-port offsets, each bin contains
\[
W_b=\left\lfloor \frac{16384}{k}\right\rfloor
\]
candidate offsets, where $W_b$ represents the bin width ($W_b=4096$ for $k=4$). Each QP is deterministically mapped to a bin using
\begin{equation}
B_i
=
I_{QP}
\bmod k,
\label{eq:bin_assign}
\end{equation}
where $I_{QP}$ denotes the QP index and $B_i$ represents the assigned source-port bin.

Within each assigned bin, the original hash-based source-port allocation mechanism is preserved. Specifically, the QP identifier is hashed using the existing Soft-RoCE hash function to generate a raw offset $o_r$, and the resulting offset is constrained to the selected bin:
\begin{equation}
o_b
=
o_r
\bmod
W_b,
\label{eq:offset_assign}
\end{equation}
where $o_r$ denotes the hash-derived raw offset, and $o_b$ denotes the offset within the assigned bin. The final source port is then generated within the QP's selected source-port range. This two-stage design introduces structural separation among QPs while preserving pseudo-random allocation within each port sub-range. Consequently, QPs communicating between the same GPU pair are less likely to generate identical packet 5-tuples, improving ECMP path diversity while preserving compatibility with commodity infrastructure. Algorithm \ref{alg:port_gen} summarizes the proposed source-port allocation procedure. 

\begin{algorithm}[H]
\DontPrintSemicolon
\SetKwFunction{FHash}{Hash32}
\SetKwInput{Input}{Input}
\Input{Queue Pair $QP$}

\BlankLine
$P_{\mathrm{base}} \leftarrow 49192$\;
$k \leftarrow 4$\;
$W_b \leftarrow 4096$\;

\BlankLine
$I_{QP} \leftarrow QP.index$\;
$o_r \leftarrow \FHash(QP.number, 14)$\;

\BlankLine
$B_i \leftarrow I_{QP} \bmod k$\;
$o_b \leftarrow o_r \bmod W_b$\;

\BlankLine
$P_s \leftarrow P_{\mathrm{base}} + (B_i \times W_b) + o_b$\;

\Return $P_s$\tcp*[r]{final assigned source port}   

\caption{Queue-Pair-Aware Source Port Allocation}
\label{alg:port_gen}
\end{algorithm}

\subsubsection{Analytical Collision Model} To provide intuition into why queue-pair-aware source-port allocation improves ECMP behavior, we analytically compare the expected ECMP path collision characteristics of the baseline and proposed mechanisms. Consider $N$ concurrent training flows (or Queue Pairs) distributed across $K$ ECMP paths. Let $p_\ell$ denote the probability that a flow is assigned to ECMP path $\ell$, where

\begin{equation}
\sum_{\ell=1}^{K} p_\ell = 1
\end{equation}

Let $h_i$ denote the ECMP path selected for flow $i$. A collision occurs when two flows are assigned to the same ECMP path. The expected number of pairwise collisions is

\begin{equation}
\mathbb{E}[C]
=
\sum_{1 \le i < j \le N}
\Pr[h_i = h_j]
\end{equation}

Since path assignments are independent under ECMP hashing,

\begin{equation}
\mathbb{E}[C]
=
\binom{N}{2}
\sum_{\ell=1}^{K} p_\ell^2
\label{eq:collision_general}
\end{equation}

Equation~(\ref{eq:collision_general}) shows that collision behavior depends directly on the ECMP path distribution.

Under ideal uniform hashing,

\begin{equation}
p_\ell=\frac{1}{K}
\end{equation}

However, distributed AI training workloads generate highly synchronized communication patterns, and correlated Queue Pair identifiers together with default source-port allocation may induce skewed ECMP path selection probabilities. Let
$\mathbf{p}^{\mathrm{base}}
=
\left(
p_1^{\mathrm{base}},
p_2^{\mathrm{base}},
\ldots,
p_K^{\mathrm{base}}
\right)$
denote the baseline ECMP path distribution. The expected collision count becomes

\begin{equation}
\mathbb{E}[C_{\mathrm{base}}]
=
\binom{N}{2}
\sum_{\ell=1}^{K}
\left(
p_\ell^{\mathrm{base}}
\right)^2
\label{eq:baseline_collision}
\end{equation}

Our QP-aware source-port allocation partitions the dynamic UDP source-port space into deterministic bins before hashing, aiming to decorrelate Queue Pair identifiers and improve ECMP path diversity.
Let
$\mathbf{p}^{\mathrm{prop}}
=
\left(
p_1^{\mathrm{prop}},
p_2^{\mathrm{prop}},
\ldots,
p_K^{\mathrm{prop}}
\right)$
denote the resulting ECMP path distribution. The expected collision count becomes
\begin{equation}
\mathbb{E}[C_{\mathrm{prop}}]
=
\binom{N}{2}
\sum_{\ell=1}^{K}
\left(
p_\ell^{\mathrm{prop}}
\right)^2
\label{eq:proposed_collision}
\end{equation}

The relative collision reduction is therefore

\begin{equation}
\Delta C 
=
1-
\frac{
\mathbb{E}[C_{\mathrm{prop}}]
}{
\mathbb{E}[C_{\mathrm{base}}]
}
\label{eq:collision_gain}
\end{equation}

Substituting Equations~(\ref{eq:baseline_collision}) and
(\ref{eq:proposed_collision}) gives

\begin{equation}
\Delta C
=
1-
\frac{
\sum_{\ell=1}^{K}
(p_\ell^{\mathrm{prop}})^2
}{
\sum_{\ell=1}^{K}
(p_\ell^{\mathrm{base}})^2
}
\end{equation}

Thus, the proposed mechanism reduces collisions whenever

\begin{equation}
\sum_{\ell=1}^{K}
(p_\ell^{\mathrm{prop}})^2
<
\sum_{\ell=1}^{K}
(p_\ell^{\mathrm{base}})^2
\label{eq:final-eq}
\end{equation}

This condition is likely to hold when the proposed binning mechanism makes the ECMP path distribution closer to uniform. Mathematically, the term $\sum_{\ell=1}^{K} p_\ell^2$ increases with path-selection skew and is minimized when all paths are selected uniformly, i.e., $p_\ell=1/K$. Since queue-pair-aware binning separates correlated QPs into different source-port regions, it reduces the likelihood that multiple QPs induce identical or highly correlated hash inputs. As a result, the induced distribution $\mathbf{p}^{\mathrm{prop}}$ is expected to be less skewed than $\mathbf{p}^{\mathrm{base}}$, leading to lower expected collisions.
Consequently, the proposed mechanism does not improve the ideal behaviour of ECMP hashing; rather, it mitigates path imbalance induced by correlated QP identifiers and synchronized distributed training traffic.

\subsection{BFD for Rapid Failure Detection}
Bidirectional Forwarding Detection (BFD) is a lightweight protocol that detects link or path failures in milliseconds, significantly faster than standard routing protocol timers \cite{rfc5880}. We propose to implement BFD with BGP in this implementation to trigger fast convergence and path switchover when a link goes down. This mechanism ensures that AI training jobs remain connected with minimal disruption during network failures.




\section{Implementation}  \label{sec:implementation}
To evaluate the building blocks described in Section \ref{sec:architecture} for geo-distributed AI training, we built a full emulated setup using open-source tools like Containerlab \cite{containerlab}, FRRouting (FRR), and Docker.
The setup models two interconnected DCs, each designed with a spine-leaf topology, enabling the deployment of geo-distributed AI training. The goal was to mimic real-world conditions involving high latency, multi tenancy, and network failures to assess the robustness and scalability of the system. 
This section details the step-by-step implementation, starting with the network topology setup, followed by device configurations and the integration of a monitoring infrastructure for observability.

\subsection{Topology Configuration and Deployment}
To simulate a geo-distributed DC environment, we used Containerlab \cite{containerlab} as the foundation for defining and deploying the network topology. Containerlab is a powerful tool that simplifies the creation of container-based network labs by allowing users to define complex topologies using a single YAML file. It abstracts away low-level container networking and lifecycle management, providing an efficient and reproducible way to emulate production-grade infrastructure.

\begin{figure}
\centering
\fbox{\includegraphics[width=\textwidth]{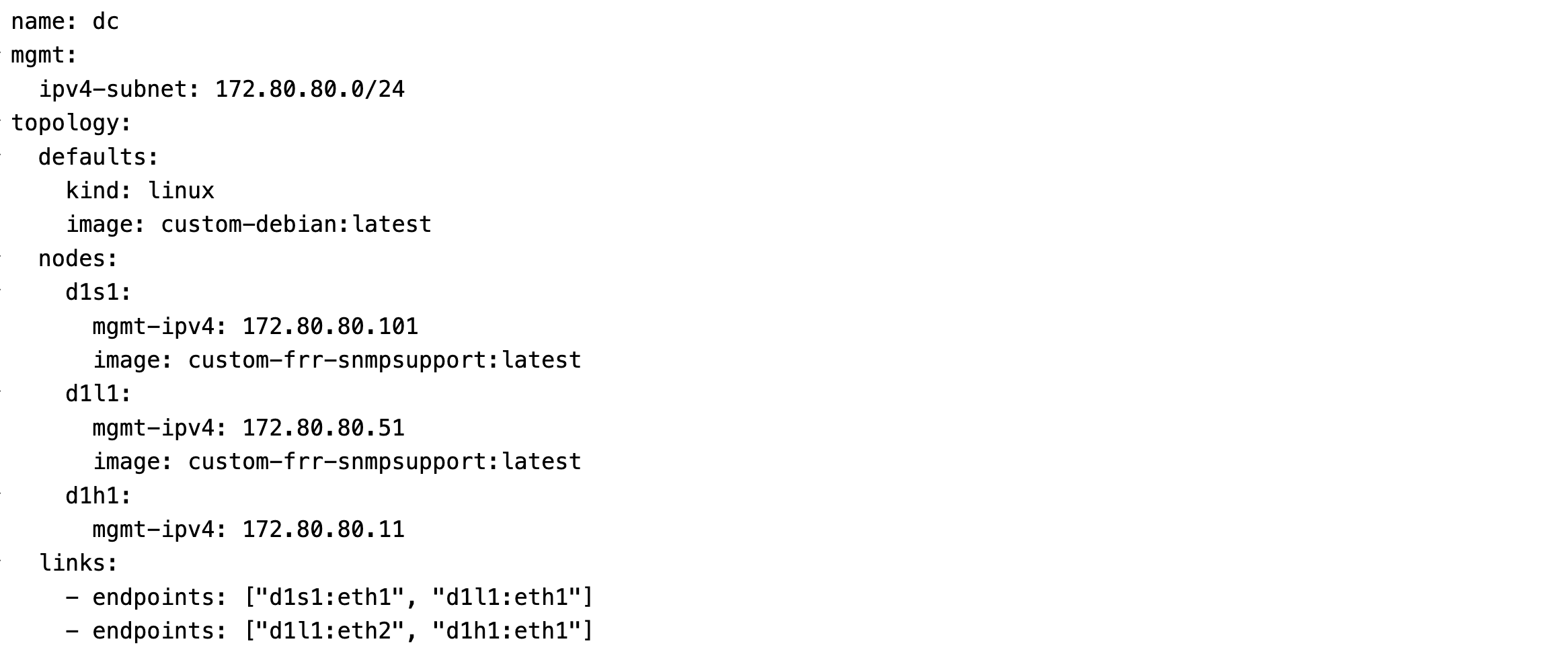}}
\caption{Example Containerlab YAML topology definition}
\label{fig:yaml_topo}
\end{figure}

The topology was defined using a YAML file that specified all the nodes, links, and bind mounts required for configuration. Each node was mapped to a container image, including FRRouting (FRR) routers for the control plane and Debian-based hosts for traffic generation and distributed training. 
Links were defined to emulate both intra-datacenter and inter-datacenter connectivity. 
Fig. \ref{fig:yaml_topo} shows an excerpt from the actual YAML topology file used in the deployment.
The emulated topology followed a spine-leaf architecture across two virtual DCs. Each data center contained two spine routers and three leaf routers. Hosts were connected to the leaf routers, and spine layers between the two data centers were linked to emulate WAN connectivity.  
Fig. \ref{fig:dist-topology} illustrates the complete topology used in the implementation. 

\begin{figure*}
\centering
\includegraphics[width=\textwidth]{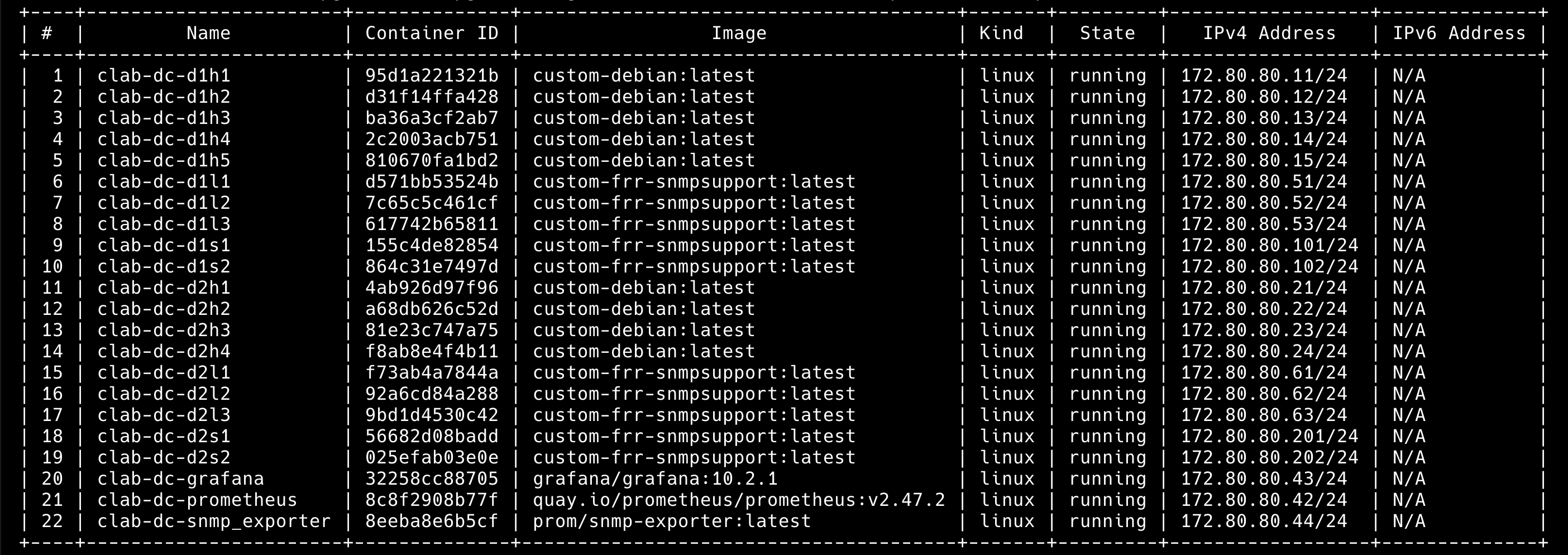}
\caption{Containerlab deployment output confirming successful startup of all nodes}
\label{fig:deployment-screenshot}
\end{figure*}

After deploying the topology, all containers were up and running, interfaces were linked, and management IPs were reachable. 
CLI screenshot confirming successful deployment is shown in \ref{fig:deployment-screenshot}. At this point, the network was ready for further configuration.

\subsection{Routers and Host Configuration}
Once the topology was deployed, the next step involved configuring routing protocols and tunnel interfaces to enable seamless Layer 2 and Layer 3 communication across the virtual data centers. 

\subsubsection{BGP EVPN Configuration}
Before deploying the topology, it was necessary to prepare the router configurations, with a particular focus on the setup required for BGP EVPN. These configurations were mounted into each router container using bind mounts defined in the Containerlab YAML file. This ensured that every router booted with its assigned routing logic and EVPN settings already in place.

Each router is configured with a unique BGP Autonomous System Number (ASN) and a router ID to identify it within the control plane. 
Peer groups are defined to simplify BGP neighbor configuration, particularly useful in spine-leaf topologies where each leaf peers with multiple spines. These peerings are established using directly connected interfaces, allowing BGP sessions to form automatically.
Within the configuration, the \texttt{address-family ipv4 unicast} block handles the exchange of traditional IP routes. The \texttt{redistribute connected} command ensures that directly connected subnets, such as loopback, are advertised into BGP.


The \texttt{address-family l2vpn evpn} block enables the EVPN control plane. Here, EVPN is activated for the defined peer group, and the \texttt{advertise-all-in} command ensures that all locally configured VXLAN Network Identifiers (VNIs) are advertised through EVPN. 
This is essential for MAC learning across the overlay, as it allows the control plane to distribute MAC-to-VTEP mappings throughout the network. Without this, hosts attached to one leaf switch would not be discoverable by hosts connected elsewhere in the fabric.

\subsubsection{VXLAN and Bridge Configuration}


Once BGP and EVPN were configured and active, the next step was to set up the VXLAN VTEPs on the leaf routers. As shown in the Fig. \ref{fig:dist-topology}, each leaf node is configured with one or more VXLAN interfaces corresponding to VNI 100, VNI 200, and VNI 300, depending on the hosts connected to it. Each VNI represents a distinct Layer 2 segment within the overlay network.



Each VXLAN interface is assigned a specific VNI, a local source IP (the VTEP IP), and is connected to a bridge that links the host-facing interface to the overlay.
This setup allows the leaf to encapsulate outgoing traffic into VXLAN format and forward it to remote VTEPs, while also decapsulating incoming traffic addressed to local hosts. It also enables proper MAC address learning and forwarding throughout the fabric via EVPN.


After VTEPs are created and connected to the bridges, they start exchanging reachability information over the EVPN control plane. Each router advertises locally known MAC addresses, VNIs, and associated VTEP IPs using BGP. This allows every router in the network to learn how to reach hosts attached to other leaves, whether in the same DC or across the WAN to a different DC. These advertisements form the basis for building and maintaining distributed Layer 2 connectivity over the Layer 3 underlay.

\subsubsection{Host Configuration}

After setting up EVPN and VXLAN, each host was manually configured with a unique MAC and IP address to maintain a consistent identity across the system. 
Once the interface is brought up, the host initiates ARP to resolve IP-to-MAC mappings. The leaf router connected to that host observes these ARP packets and learns the MAC-to-IP binding. It then advertises this information to the rest of the fabric using EVPN. As a result, all routers in the overlay learn how to reach every host, making inter-host communication seamless across the distributed data centers.


\subsubsection{ECMP and BFD Configuration}

\begin{figure*}s
    \centering
    \includegraphics[width=0.7\textwidth]{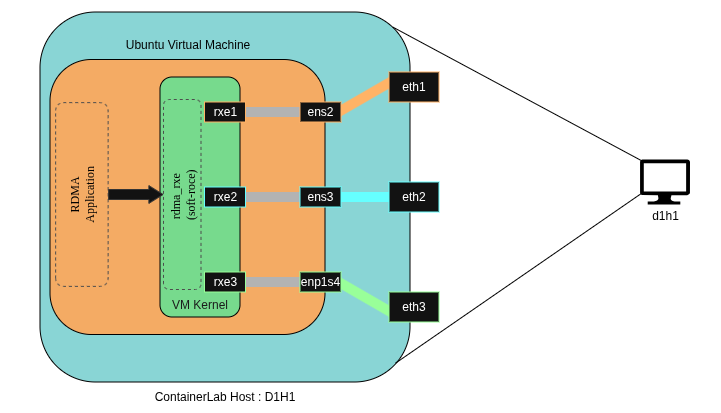} 
    \caption{Container-based Soft-RoCE VM implementation.}
    \label{fig:vm-implementation}
\end{figure*}

To enhance both fault tolerance and traffic distribution, ECMP and BFD were configured. 
Algorithm \ref{alg:port_gen} was implemented in ContainerLab by modifying the Soft-RoCE module \cite{rxe-git}, a software implementation of RDMA communication. 
To enable per-node RDMA isolation within the emulated environment, each host was instantiated as a lightweight QEMU virtual machine inside a Docker container, providing an independent guest kernel per emulated node. 
As illustrated in Fig.~\ref{fig:vm-implementation}, separate \texttt{rdma-rxe} driver instances were maintained for individual network connections, each associated with an independent VLAN and QP domain. This design preserves isolated RDMA state per node, including QPs, RDMA endpoint addressing state, and connection management structures.

To integrate the virtual machines with the ContainerLab topology, VM network interfaces were bridged through QEMU tap interfaces into the Docker container network namespace and subsequently connected to external \texttt{veth} interfaces attached to the emulated fabric. Consequently, RoCEv2 traffic generated within guest virtual machines traverses the emulated multi-hop data center network while preserving RDMA isolation, allowing each node to operate as an independent RDMA-capable endpoint. Alongside ECMP, BFD was configured to enable fast failure detection between BGP peers.


\subsubsection{Control-plane and data-plane operations}

\begin{figure*}
    \centering
    \includegraphics[width=\linewidth]{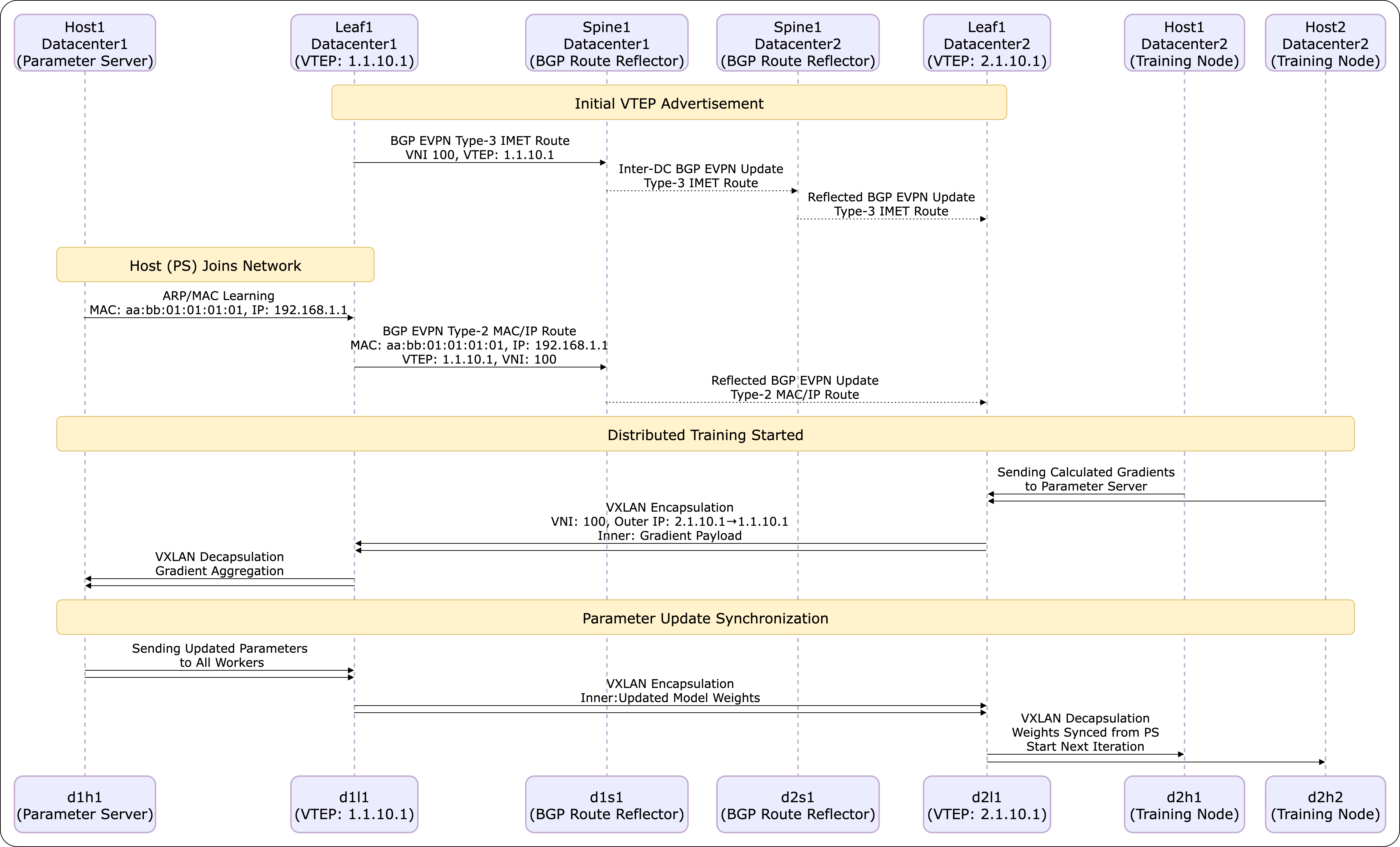}
    \caption{VXLAN+EVPN control/data plane operations: model training using Parameter Server (PS) architecture in geo-distributed AI data centers.}
    \label{fig:evpn_flow}
\end{figure*}

Fig.~\ref{fig:evpn_flow} illustrates the VXLAN–EVPN control- and data-plane operations that enable distributed machine-learning training across the two-datacenter topology shown in Fig.~\ref{fig:dist-topology}. The sequence shows how the fabric provides seamless communication between a parameter server in one datacenter and training nodes in the other. For clarity, the sequence diagram includes one parameter server (\texttt{d1h1}) and only two training nodes (\texttt{d2h1} and \texttt{d2h2}) connected through VNI 100 in Fig.~\ref{fig:dist-topology}.

The initialization phase begins with VTEP advertisement, where leaf switch \texttt{d1l1} (Leaf1 of Datacenter1) sends an EVPN Type-3 Inclusive Multicast Ethernet Tag (IMET) route to its local route reflector \texttt{d1s1} (Spine1 of Datacenter1), advertising its VTEP address (1.1.10.1) for \texttt{VNI 100}. This advertisement is propagated through inter-datacenter BGP peering to \texttt{d2s1} (Spine1 of Datacenter2) and subsequently reflected to \texttt{d2l1} (Leaf1 of Datacenter2), establishing the foundation for cross-datacenter VXLAN connectivity.

As the parameter server \texttt{d1h1} becomes active in Datacenter1, it sends an ARP request, allowing \texttt{d1l1} to learn its MAC address. \texttt{d1l1} then generates an EVPN Type-2 MAC/IP advertisement containing the host's identity (\texttt{MAC: aa:bb:01:01:01:01, IP: 192.168.1.1}) and its VTEP association (1.1.10.1). This advertisement is distributed via the BGP route reflector hierarchy to \texttt{d2l1}, enabling Datacenter2 to establish reachability to the parameter server.

During the distributed training phase, training nodes (\texttt{d2h1} and \texttt{d2h2}) send their calculated gradients to the parameter server. \texttt{d2l1} encapsulates these gradient payloads using VXLAN with the learned VTEP information and forwards them across the inter-datacenter fabric to \texttt{d1l1}, which decapsulates and delivers the gradients to \texttt{d1h1} for aggregation.

Following gradient processing, \texttt{d1h1} initiates parameter synchronization by sending the updated model weights to all training workers. \texttt{d1l1} encapsulates these parameter updates using VXLAN and forwards them across the inter-datacenter fabric to \texttt{d2l1}. Upon receiving the encapsulated traffic, \texttt{d2l1} performs VXLAN decapsulation and delivers the synchronized weights to the training nodes, completing the parameter update cycle and preparing them for the next iteration.

\subsection{Monitoring Infrastructure}

\begin{figure*}
\centering
\fbox{\includegraphics[width=\textwidth]{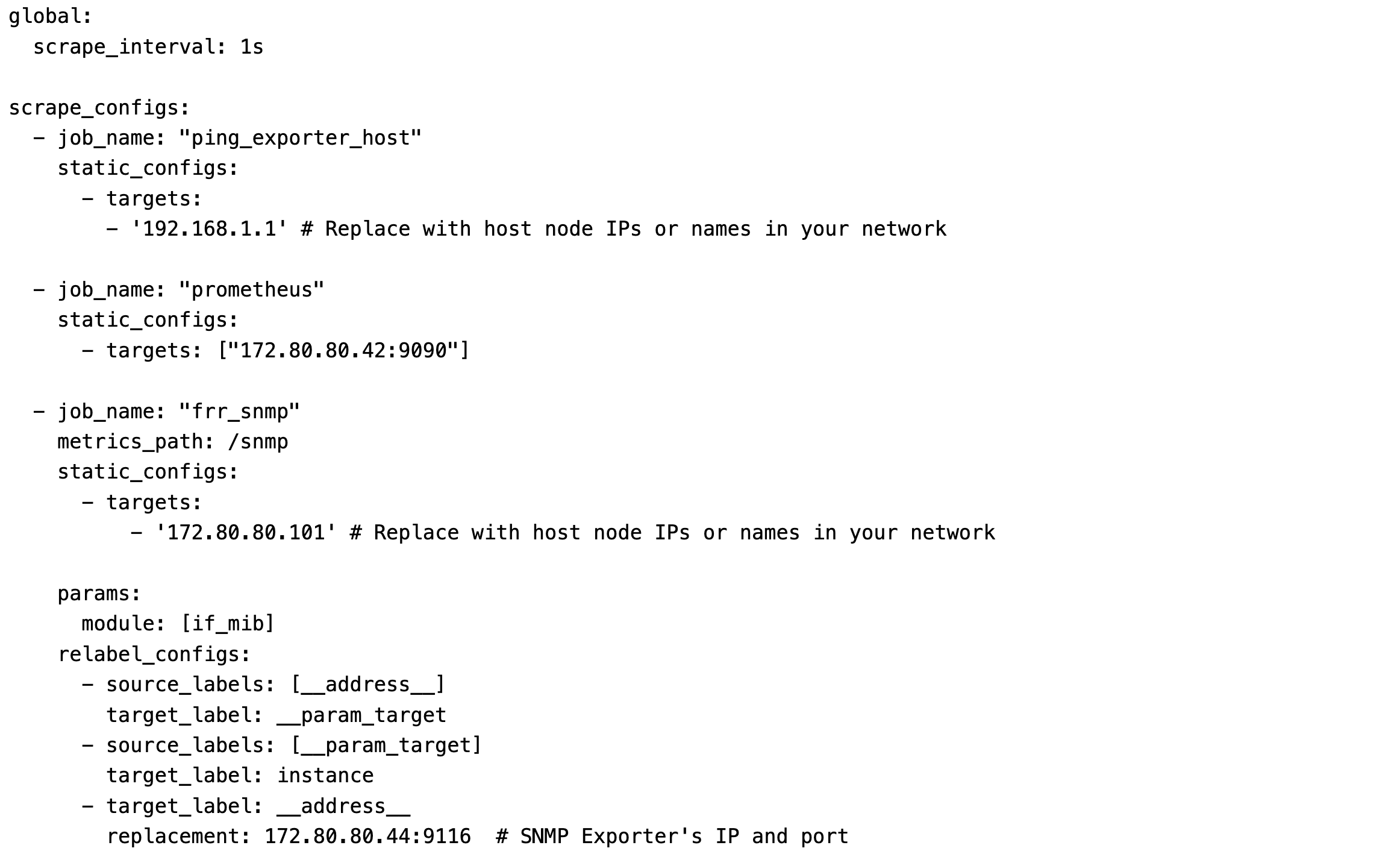}}
\caption{Prometheus configuration file with SNMP and Ping Exporter targets}
\label{fig:prometheus_config}
\end{figure*}

To ensure visibility into the health and performance of the emulated geo-distributed network, a monitoring stack was deployed using open-source tools. 
Prometheus was used as the central monitoring system, with SNMP and Ping Exporters installed on various nodes to expose metrics related to interface statistics, round-trip time, and system availability. 
This setup enabled real-time tracking of network conditions and protocol behavior during training experiments.
Prometheus was configured using a YAML file that defined monitoring targets, including the IP addresses of routers and hosts. It periodically scraped metrics from each target using HTTP endpoints exposed by the exporters. 
Fig. \ref{fig:prometheus_config} shows a portion of the Prometheus configuration file that specifies SNMP and Ping Exporter endpoints.
\begin{figure*}
\centering
\fbox{\includegraphics[width=\textwidth]{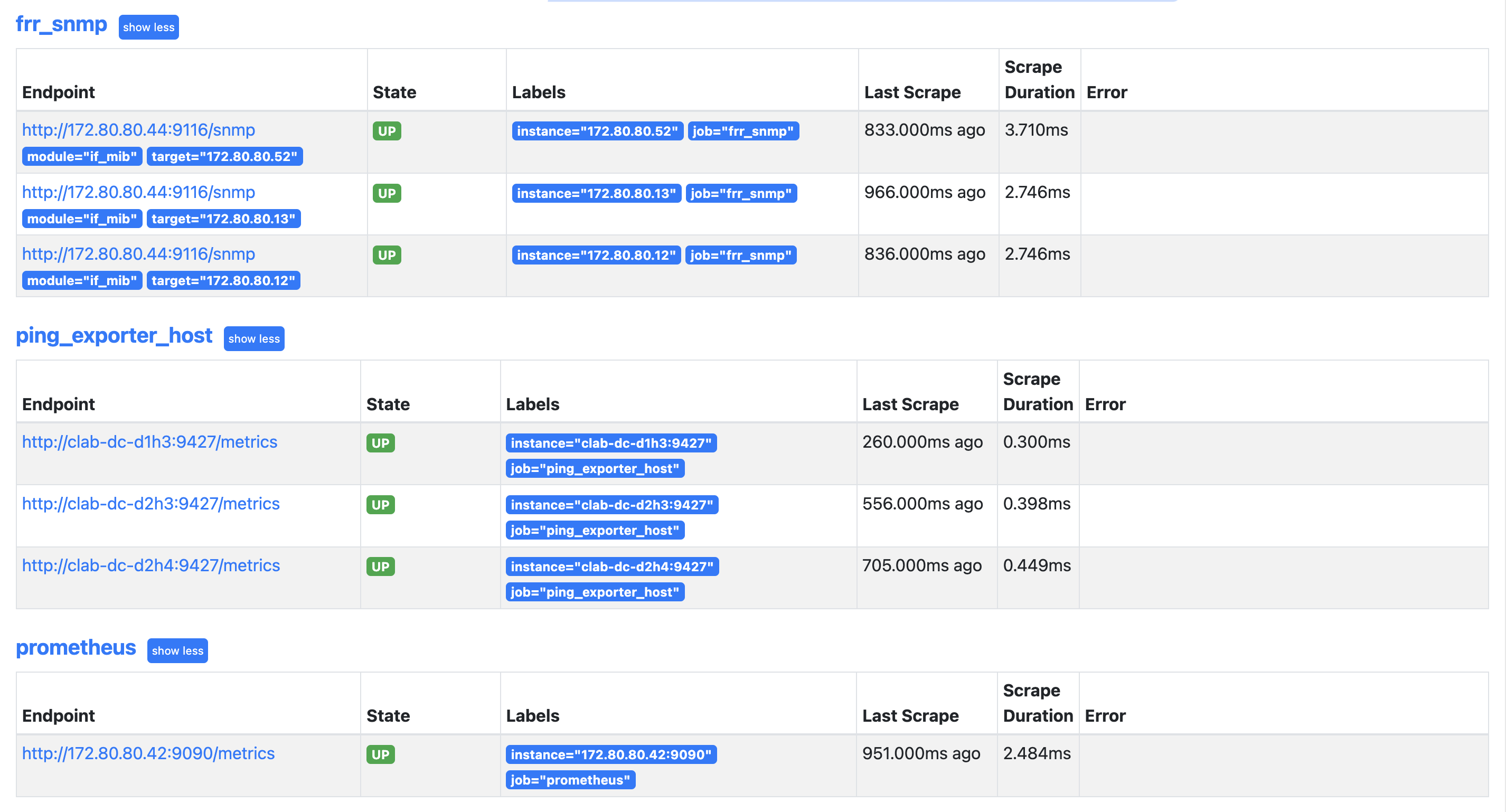}
}
\caption{Prometheus showing active monitoring targets}
\label{fig:prometheus_targets}
\end{figure*}
Once configured, the Prometheus web interface confirmed that all targets were reachable and actively exporting metrics. 
Fig. \ref{fig:prometheus_targets} shows the Prometheus Targets page, where each entry corresponds to a monitored router or host in the topology. A green status indicates successful metric collection, confirming that the exporters were running correctly.

This monitoring setup provided key insights during the distributed training experiments, such as link availability, latency between data centers, interface usage, and protocol stability. By combining these metrics with packet captures and protocol logs, it was possible to correlate training performance with underlying network behavior. This observability was crucial in validating the impact of failure recovery mechanisms, ECMP behavior, and multi-tenant isolation in the experimental evaluations.

\section{Experiments}  \label{sec:experiments}
This section presents the experimental evaluation of the proposed geo-distributed AI training system. The experiments are structured to address key aspects of network performance and distributed training, including reachability, ECMP behavior, link failure recovery, multi-tenancy, and model training. 
All experiments were conducted on the emulated spine-leaf topology described in the previous section.

\subsection {Reachability (Ping Test)}
To verify end-to-end connectivity and establish baseline network performance, ICMP echo requests were sent from all hosts to all other hosts in the emulated topology. 



To emulate the latency and jitter typically observed in real-world geo-distributed data centers, artificial delay and jitter were introduced on each inter-datacenter link using Containerlab's \texttt{netem} tool. 
Specifically, a fixed delay of 5 ms and a jitter of 1 ms were applied per link. 

\begin{figure}
    \centering
    \includegraphics[width=0.8\linewidth]{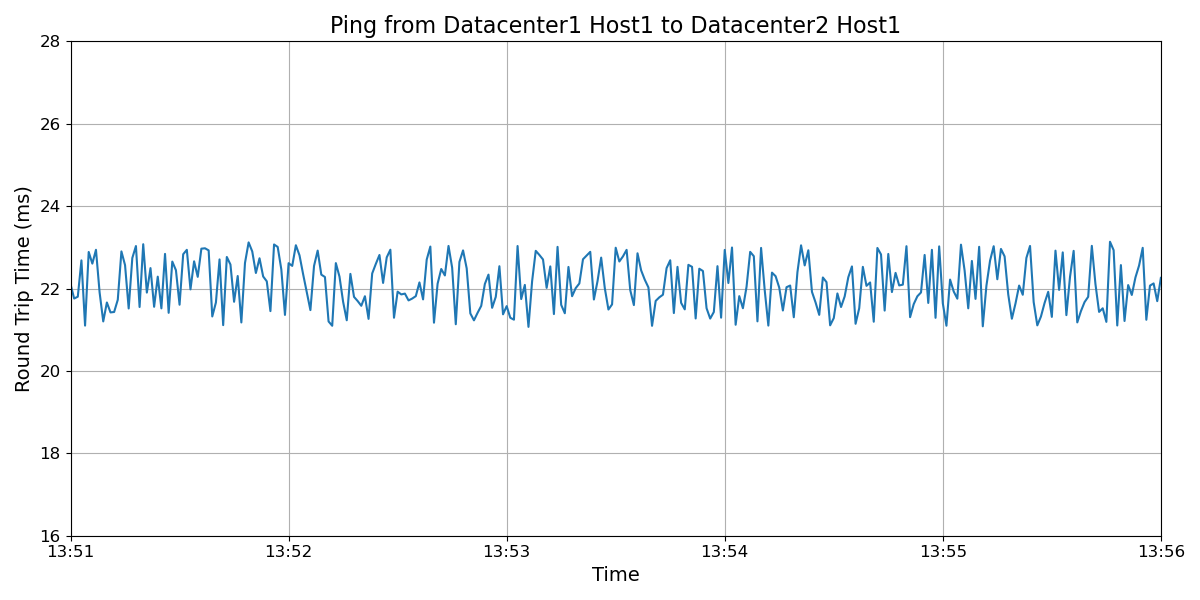}
    \caption{Ping RTT from Datacenter1 Host1 to Datacenter2 Host1 after introducing artificial delay and jitter using Containerlab's netem tool.}
    \label{fig:ping-netem}
\end{figure}

Fig.~\ref{fig:ping-netem} shows the round-trip time (RTT) measured when pinging from Datacenter1 Host1 to Datacenter2 Host1.  As shown in Fig. ~\ref{fig:ping-netem}, the RTT is approximately 22 ms, with observed variation consistent with the configured jitter. This experiment ensures that the emulated network environment closely resembles the latency and variability of production geo-distributed AI training.


\subsection{Equal-Cost Multi-Path Routing (ECMP)}
To evaluate the ECMP routing, multiple traffic flows were generated from \texttt{d1h1} to \texttt{d2h2}.
As shown in the topology diagram (Fig. \ref{fig:dist-topology}), the goal was to observe how the network distributes traffic across parallel paths to improve throughput and resilience.

\begin{figure*}
    \centering
    \includegraphics[width=\textwidth]{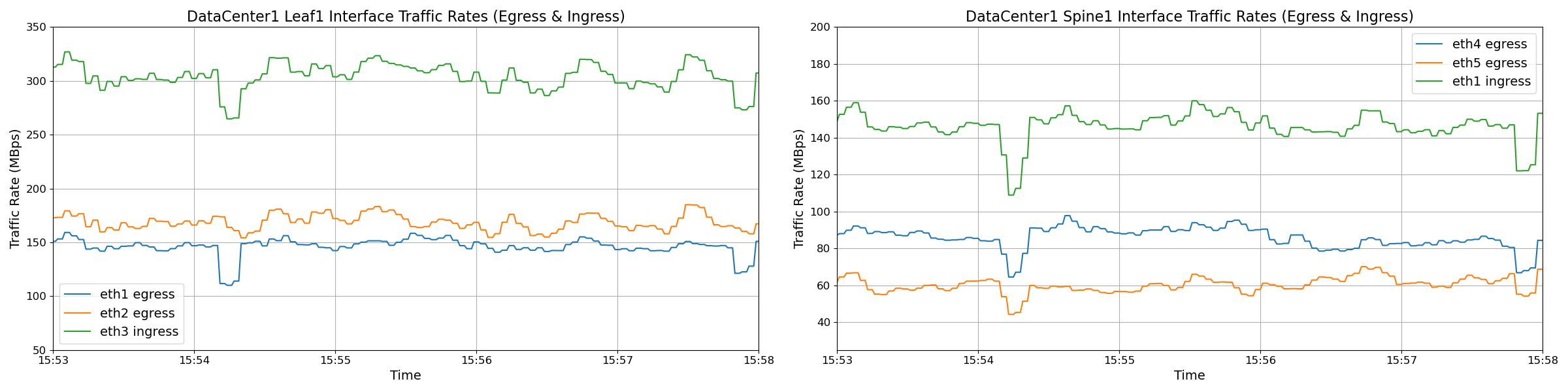}
    \caption{\textbf{[Left]}: At the leaf node (Leaf1, DC1), traffic is evenly distributed across both uplinks to the spine layer. \textbf{[Right]}: At the spine node (Spine1, DC1), traffic is further balanced across two WAN links toward DC2, confirming ECMP functionality at both the aggregation and core layers.}
    \label{fig:ecmp-demonstration}
\end{figure*}

At leaf node \texttt{d1l1}, the traffic from the host interface (eth3) is divided into two uplinks (eth1 and eth2), which connect to a different spine switch within the DC. This shows that ECMP effectively distributes the outgoing packets on both available spine paths, as shown in Fig.~\ref{fig:ecmp-demonstration} (left). 
On the spine node \texttt{d1s1}, the incoming traffic from the leaf interface (eth1) is further distributed across two WAN-facing interfaces (eth4 and eth5), each leading to a different spine in the remote data center. This confirms that ECMP continues to balance flows even at the spine layer, ensuring optimal utilization of all inter-datacenter links, as shown in Fig.~\ref{fig:ecmp-demonstration} (right).





\begin{figure}
\centering
\includegraphics[width=0.8\textwidth]{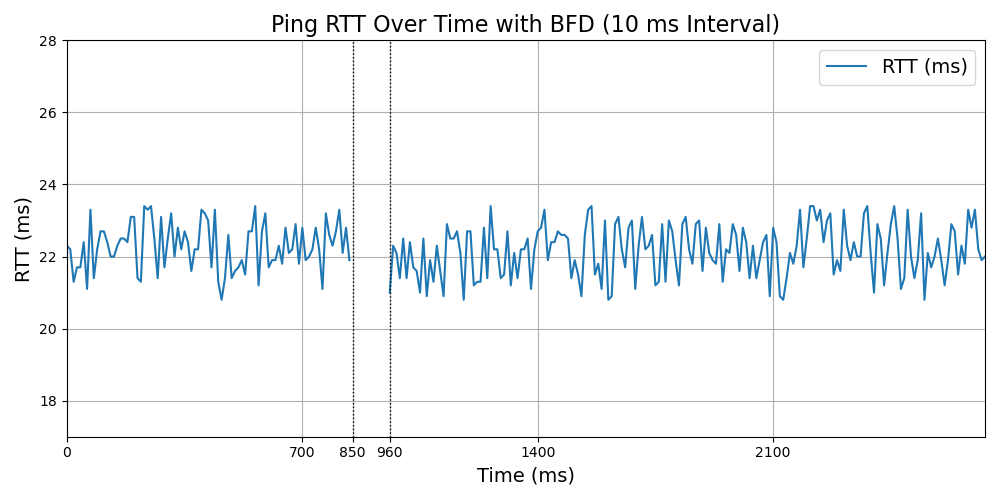}
\caption{Ping RTT over time with BFD enabled (10 ms interval, 3 retries). Recovery is achieved in approximately 110 ms.}
\label{fig:link-failure-bfd10}
\end{figure}



We also evaluated the proposed source port allocation strategy.
To identify an effective bin partitioning strategy, we first experimented with multiple bin configurations. Our preliminary analysis showed that partitioning the dynamic UDP source port range into 4 bins provided the most stable and consistent improvement in traffic distribution compared to other configurations. Based on this observation, we fixed the number of bins to 4 and conducted experiments to validate the proposed approach under varying levels of communication parallelism. Specifically, we varied the number of internal channels (QPs) across 4, 8, 16, and 32, to assess the effectiveness of the strategy using Eq.(\ref{conga-formulation}) \cite{alizadeh2014conga}:

\begin{equation}
    \text{Load Factor} = \frac{U_{max} - U_{min}}{U_{avg}},
    \label{conga-formulation}
\end{equation}

where $U_{max}$, $U_{min}$, $U_{avg}$ represent the maximum, minimum, and average amount of data (bytes) transmitted over the actively utilized links. To avoid distortion in the metric when the number of flows is smaller than the available ECMP paths, only links carrying traffic above a predefined threshold are considered in the calculation. Specifically, a link is classified as \textit{used} if the total transmitted bytes exceed the threshold. The values of $U_{max}$, $U_{min}$, $U_{avg}$ are then computed only across these active links, preventing \textit{underutilized} or \textit{idle} links from artificially lowering the ratio.

    \begin{figure}
        \centering
        \centering
        \includegraphics[width=0.8\textwidth]{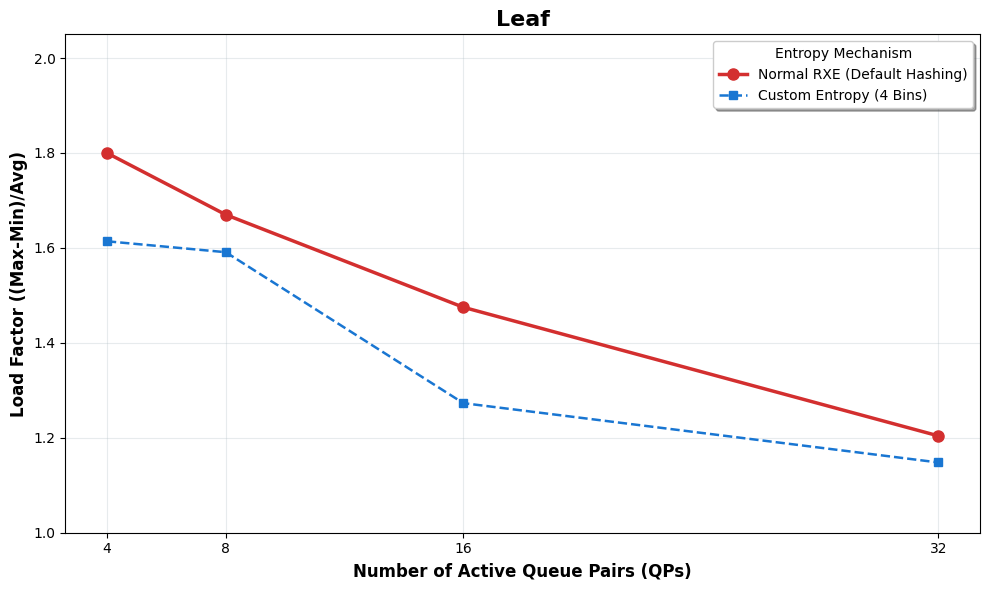}
        \caption{Load Factor at Leaf Switch}
        \label{fig:Leaf Ratio}
    \end{figure}
    
    \begin{figure}
        \centering
        \includegraphics[width=0.8\textwidth]{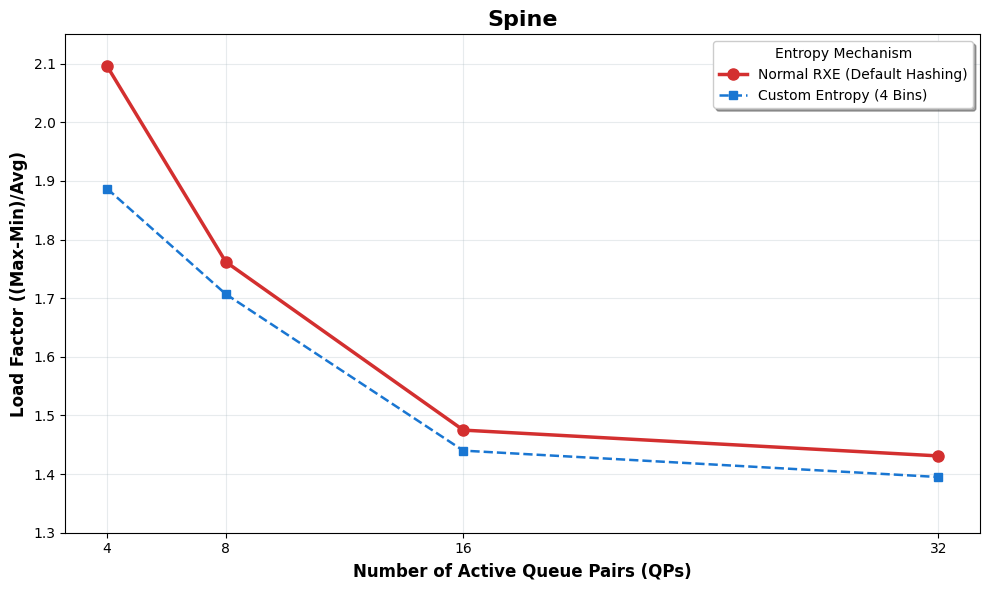}
        \caption{Load Factor at Spine Switch}
        \label{fig:Spine Ratio}
    \end{figure}
Fig. \ref{fig:Leaf Ratio} and \ref{fig:Spine Ratio} compare the load factor of the default RXE hashing mechanism and the proposed 4-bin source port allocation strategy across leaf and spine nodes under different numbers of QPs.
The proposed 4-bin source port allocation strategy demonstrates a significant improvement compared to the default RXE allocation. 
At the leaf node, we achieve a peak improvement of 13.7\% at 16 QPs, whereas at the spine node, the maximum improvement of 9.9\% is observed at 4 QPs.
The proposed strategy reduces the likelihood of allocating the same source port number, thereby lowering the probability of ECMP hash collisions. The performance gain gradually converges as the number of QPs increases due to the higher natural flow entropy introduced by additional communication channels.


\subsection{Link Failure Recovery}

To evaluate the network's resilience and convergence time during failures, we conducted a link failure recovery experiment under realistic WAN conditions. The test involved transferring data between two hosts with the configured inter-datacenter round-trip latency of 22 ms and 1 ms jitter using Containerlab's \texttt{netem} tool. During continuous ping measurements, a critical path was disrupted by dropping all packets on the route, effectively emulating a link failure.

\begin{figure}
\centering
\includegraphics[width=0.8\textwidth]{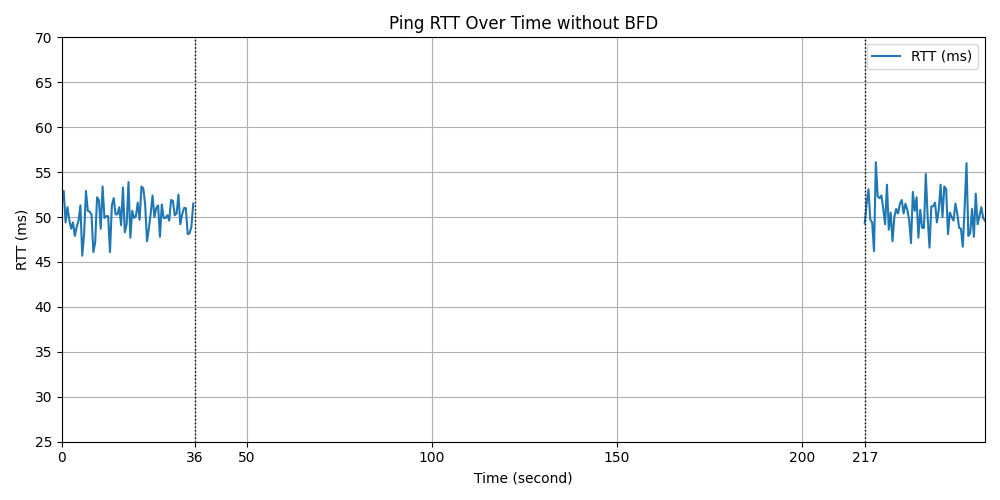}
\caption{Ping RTT over time during link failure recovery using default BGP timers. Recovery takes approximately 180 seconds due to slow failure detection.}
\label{fig:link-failure-nobfd}
\end{figure}

When only default BGP timers were used (keepalive interval of 60 seconds and hold timer of 180 seconds), the network required the full hold timer duration to detect the failure and converge to a new path. 
As shown in ~\ref{fig:link-failure-nobfd}, the RTT remains disrupted for approximately 180 seconds following the failure event. 
This prolonged convergence time is unsuitable for distributed training, as it can lead to significant computation loss and degraded performance.


To enable rapid failure detection, BFD was configured between BGP peers. 
With BFD enabled at 10 ms with 3 retries, the network detected the failure and restored connectivity in approximately 110 ms, as depicted in Fig. \ref{fig:link-failure-bfd10}. 
This quick failover is essential for minimizing disruption during AI training.
These results demonstrate that integrating BFD with BGP-EVPN overlays dramatically reduces network convergence time following a link failure-from several minutes with default BGP timers to well under a second with aggressive BFD settings. This rapid failover capability is vital for sustaining high-performance and resilient distributed DNN training across geo-distributed data centers.



\subsection{Multi-Tenancy}
In our setup, multi-tenancy is implemented using VNIs, with each training workload assigned a unique VNI. 
This ensures that traffic from different jobs remains logically isolated at the overlay level, independent of the underlying physical topology. The experimental setup, based on the spine-leaf topology described in Fig.~\ref{fig:dist-topology}, was configured with multiple VNIs to emulate separate tenant environments. Hosts were assigned to different VNIs, and connectivity tests were conducted to verify both intra-tenant communication and inter-tenant isolation.

The results of the multitenancy experiment are summarized in Table \ref{tab:multitenancy}. 
Hosts within the same VNI could communicate successfully, as shown by the low round-trip times between \texttt{d1h1} and \texttt{d2h1} (both in \texttt{VNI 100}), and between \texttt{d1h3} and \texttt{d1h5} (both in \texttt{VNI 200}). 
In contrast, attempts to communicate across VNIs, such as from \texttt{d1h2} in \texttt{VNI 100} to \texttt{d1h3} in \texttt{VNI 200}, or from \texttt{d1h4} in \texttt{VNI 300} to \texttt{d2h4} in \texttt{VNI 200}, resulted in \texttt{destination host unreachable}, confirming that the proposed system enforces strict tenant isolation, essential for supporting concurrent AI training workloads.




These results confirm that the system strictly enforces multitenancy, allowing only intra-VNI communication while blocking inter-VNI traffic. This isolation is critical for concurrently running multiple AI training workloads, ensuring privacy, performance isolation, and security across tenants in a shared infrastructure.

\begin{table}
\small
\caption{Ping results within and across VXLAN segments.}
\resizebox{\columnwidth}{!}{\begin{tabular}{|c|c|c|c|c|}
\hline
\textbf{Source Host} & \textbf{Source VNI} & \textbf{Destination Host} & \textbf{Destination VNI} & \textbf{Ping time} \\
\hline
d1h1 & 100 & d2h1 & 100 & 21.4 ms \\
d1h3 & 200 & d1h5 & 200 & 0.07 ms \\
d1h2 & 100 & d1h3 & 200 & destination host unreachable \\
d1h4 & 300 & d2h4 & 100 & destination host unreachable\\

\hline
\end{tabular}}
\label{tab:multitenancy}
\end{table}


\subsection{Distributed Training}
To evaluate the multi-tenancy and network performance of distributed deep learning workloads, we conducted simultaneous training using both AllReduce and Parameter Server (PS) architectures over the emulated spine-leaf topology shown in Fig.~\ref{fig:dist-topology}. Both training jobs fine-tuned the DistilGPT2 model (approximately 82 million parameters) on the WikiText-2 dataset. The AllReduce job, implemented using PyTorch DistributedDataParallel (DDP), operated on VNI 300, while the PS job, implemented using Ray and PyTorch, ran on VNI 100, ensuring network-level isolation across the virtual fabric.

\begin{figure}
\centering
\includegraphics[width=0.8\textwidth]{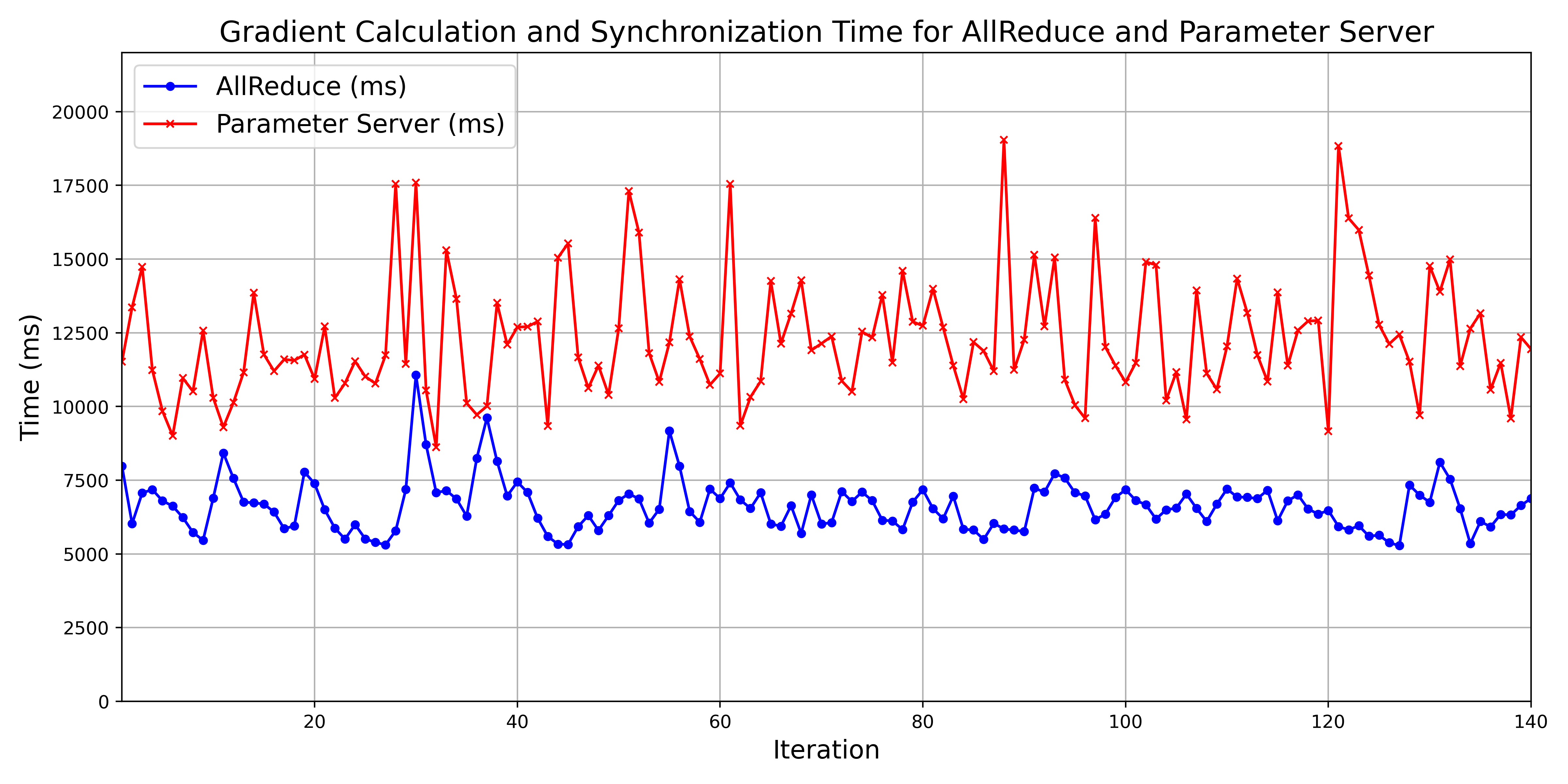}
\caption{Batch-wise gradient computation and synchronization time for Allreduce and Parameter Server architecture.}
\label{fig:sync-time}
\end{figure}

The effective throughput over the spine links was approximately 800~Mbit/s, constrained by the configured 22~ms round-trip latency, thereby introducing a realistic communication bottleneck \cite{nvidia2024turbocharge}. In the AllReduce setup, four worker nodes participated in collective gradient synchronization, with each batch generating approximately 312~MB of gradient data. As shown in Fig.~\ref{fig:sync-time}, the per-batch training time varied between 5{,}000~ms and 11{,}000~ms.

\begin{table*}[t]
\centering
\caption{Inter-data-center overlay approaches relevant to geo-distributed AI training}
\label{tab:overlay-dci-comparison}

\resizebox{\textwidth}{!}{%
\begin{tabular}{|p{3cm}|p{3.3cm}|p{3.3cm}|p{3.3cm}|}
\hline

\textbf{Feature / Requirement}
&
\textbf{VXLAN + EVPN}
&
\textbf{IP over GRE/MPLS}
&
\textbf{SD-WAN / VPLS \cite{vpls-survey}}
\\
\hline

Primary Role
&
Scalable L2/L3 overlay and multi-tenant inter-DC connectivity
&
WAN transport and L3 tunneling
&
Enterprise WAN connectivity and virtualization
\\
\hline

Layer-2 Extension
&
Native L2 over L3 via VXLAN
&
Requires additional overlay mechanisms
&
Supported (VPLS); SD-WAN varies by implementation
\\
\hline

Tenant Isolation / Virtualization
&
High (VXLAN VNIs + EVPN control plane)
&
Requires VRFs / overlay mechanisms
&
Moderate; platform dependent
\\
\hline

Control Plane
&
Dynamic (MP-BGP EVPN)
&
MPLS control protocols / static GRE
&
Controller-driven or policy-based
\\
\hline

Endpoint Discovery / Reachability
&
Automated
&
Manual / protocol dependent
&
Controller dependent
\\
\hline

Operational Scalability
&
Designed for large-scale multi-tenant environments
&
Depends on deployment architecture
&
Platform dependent
\\
\hline

Resilience Enhancements
&
Can integrate fast failure detection (e.g., BFD)
&
Depends on routing design
&
Controller dependent
\\
\hline

AI Infrastructure Applicability
&
Promising for geo-distributed AI overlays
&
Provides baseline inter-site transport
&
Primarily enterprise-focused
\\
\hline

\end{tabular}
}
\end{table*}

In contrast, the PS setup consisted of one parameter server and four workers, with each batch producing approximately 459 MB of gradient data. The training time in this case ranged from 9{,}000 ms to 18{,}000 ms, as shown in Fig.~\ref{fig:sync-time}, due to the higher communication volume and centralized aggregation.

The results indicate that AllReduce achieved lower and more consistent per-batch training times compared to the Parameter Server setup, despite involving decentralized gradient synchronization. The PS architecture, while centralized, exhibited higher variance and longer batch durations, likely due to bandwidth saturation and contention at the server node.

\section{Related Work}  \label{sec:related-works}

Research related to this work spans geo-distributed AI training, inter--data-center overlays, and transport fabrics for large-scale distributed AI systems.

\subsection{Distributed and Geo-Distributed AI Training}

Early work on distributed deep learning focused on scaling training within a single data center using data parallelism, parameter servers \cite{li2014communication}, and optimized collective communication \cite{romero2022accelerating}. More recent systems support large-scale AI training using hybrid and hierarchical parallelism but typically assume tightly coupled, low-latency intra-data-center environments \cite{ByteScale}. Geo-distributed and collaborative learning have been explored to improve data locality, privacy preservation, and fault tolerance \cite{song2024hcec, L3DML, li2025parallel}. Federated and communication-efficient learning techniques further reduce synchronization overhead by limiting communication frequency or exchanged information \cite{liang2024communication}. Collaborative training across multiple data centers has also been studied to improve scalability and regulatory compliance \cite{xu2018collaborativedeeplearningmultiple, mi2020collaborative}. However, these efforts largely abstract away the underlying network infrastructure and do not investigate how inter-data-center fabrics should be designed to support synchronization-intensive distributed AI workloads.

\subsection{Inter--DC Overlays}

Overlay-based network virtualization has become the foundation of modern cloud and multi--data-center deployments. VXLAN enables scalable Layer-2 overlays over Layer-3 infrastructure \cite{cai2013evolution,rfc7348}, while EVPN provides a standards-based control plane supporting MAC/IP reachability, endpoint discovery, and Layer-2/Layer-3 integration \cite{rfc8365}. These technologies are widely adopted because of their scalability, operational flexibility, and support for workload mobility across geographically distributed infrastructure \cite{8459946}.

Prior studies on EVPN and VXLAN have primarily focused on cloud networking, service-oriented environments, and general-purpose data-center interconnection \cite{10629438,8459946}. To the best of our knowledge, limited prior work has systematically investigated EVPN and VXLAN specifically from the perspective of geo-distributed AI training workloads. Distributed AI training introduces unique communication behavior, including bursty synchronized collective traffic, long-lived elephant flows, and sensitivity to transient failures and path imbalance. These workload characteristics motivate revisiting existing overlay mechanisms in the context of AI infrastructure.

Legacy interconnection approaches such as GRE/MPLS and enterprise-focused overlays such as SD-WAN or VPLS provide baseline WAN connectivity but offer limited support for large-scale tenant isolation, operational flexibility, and communication patterns associated with synchronization-intensive AI workloads. In contrast, VXLAN+EVPN combines scalable Layer-2 virtualization with standards-based control-plane automation, providing a promising foundation for geo-distributed AI infrastructure. Our work investigates this applicability while introducing enhancements targeting resilience and communication efficiency in distributed AI environments. 
Table \ref{tab:overlay-dci-comparison} summarizes the inter-DC overlay approaches and their features.

\subsection{Transport Fabrics for Distributed AI Workloads}

Transport fabrics determine how training traffic is efficiently carried within AI clusters, while overlay technologies determine how geographically distributed infrastructure is interconnected and managed. Three transport technologies are particularly relevant to our work.

RoCEv2 extends RDMA semantics over Ethernet/IP, providing low latency and CPU bypass while relying on specialized NICs and carefully engineered congestion-control mechanisms \cite{rdma-meta, rdma-hyperscale}. Ultra Ethernet (UEC) \cite{hoefler2025ultra} is a recent Ethernet-native transport designed for AI and HPC environments, emphasizing extremely high throughput and explicit congestion management. InfiniBand with SHARP \cite{rdma-hyperscale} represents the performance ceiling for AI collectives through hardware-supported lossless communication and in-network reduction.

These technologies primarily target tightly controlled intra-data-center deployments. Geo-distributed AI training introduces additional requirements beyond transport efficiency, including inter-data-center connectivity, tenant isolation, workload mobility, and resilient operation across geographically dispersed sites. Our work focuses primarily on this interconnection layer, investigating how VXLAN+EVPN can complement transport fabrics by providing scalable, multi-tenant, and resilient overlay infrastructure for geo-distributed AI training.

\subsection{Emulation and Evaluation Frameworks}

Large-scale geo-distributed AI infrastructure is expensive to deploy and difficult to evaluate under controlled conditions. Network emulation frameworks provide practical mechanisms for evaluating networking protocols and distributed systems before production deployment. Container-based network emulation enables reproducible experimentation with realistic protocol stacks \cite{handigol2012reproducible, containerlab}, 
while simulators such as ns-3 support controlled evaluation of protocol behavior \cite{ns-3}.
In parallel, ML system simulators such as ASTRA-sim enable design-space exploration for large-model training under different communication and topology assumptions \cite{won2023astra}.
Existing platforms provide general-purpose protocol validation but do not explicitly target geo-distributed AI networking environments involving synchronized collective communication, WAN overlays, and AI-oriented traffic patterns. This work complements these efforts by providing an evaluation framework tailored to emerging geo-distributed AI training infrastructure.

\section{Conclusion and Future Work}
\label{sec:conclusion}

This paper investigated EVPN--VXLAN as an inter-data-center networking substrate for geo-distributed AI training environments. We presented a practical and extensible emulation framework that enables systematic evaluation of distributed AI training over geographically distributed data centers while incorporating realistic overlay networking and WAN conditions. In addition, we introduced network-level enhancements aimed at improving communication efficiency and resilience for synchronization-intensive AI workloads.
Built using open-source tools including ContainerLab and FRRouting (FRR), the proposed framework enables reproducible experimentation and supports the joint exploration of AI training systems and network infrastructure design. The presented results provide initial insights into communication behavior, failure recovery, and traffic distribution mechanisms in geo-distributed AI environments.
Future work includes evaluating larger training workloads, studying additional orchestration environments, and exploring adaptive routing and traffic engineering mechanisms for improving communication efficiency in dynamic multi-data-center deployments.




\printcredits

\bibliographystyle{cas-model2-names}

\bibliography{references}



\end{document}